\documentclass{iaus}
\usepackage{graphicx}

\title[Cold H {\footnotesize I} in Turbulent Eddies and Galactic Spiral Shocks] 
{Cold H {\Large I} in Turbulent Eddies and Galactic Spiral Shocks}

\author[S. J. Gibson et al.]   
{
Steven J. Gibson$^1$,
A. Russell Taylor$^2$,
Jeroen M. Stil$^2$,
Christopher
\break\
M. Brunt$^3$,
Dain W. Kavars$^4$,
\and 
John M. Dickey$^{4,5}$
}

\affiliation{
$^1$Arecibo Observatory, National Astronomy and Ionosphere Center, Arecibo, PR 00612, U.S.A.\\[\affilskip]
$^2$Dept. of Physics \& Astronomy, University of Calgary, Calgary, Alberta T2N~1N4, Canada\\[\affilskip]
$^3$School of Physics, University of Exeter, Exeter, United Kingdom EX4 4QL\\[\affilskip]
$^4$Department of Astronomy, University of Minnesota, Minneapolis, MN 55455, U.S.A.\\[\affilskip]
$^5$School of Mathematics and Physics, University of Tasmania, 
Hobart, TAS 7001, Australia
}

\pubyear{2007}
\volume{237}  
\pagerange{119--126}
\date{?? and in revised form ??}
\jname{TRIGGERED STAR FORMATION IN A TURBULENT ISM}
\editors{A.C. Editor, B.D. Editor \& C.E. Editor, eds.}
\begin{document}

\maketitle

\begin{abstract}
{\sc H i} 21cm-line self-absorption (HISA) reveals the shape and distribution
of cold atomic clouds in the Galactic disk.  Many of these clouds lack
corresponding CO emission, despite being colder than purely atomic gas in
equilibrium models.  HISA requires background line emission at the same
velocity, hence mechanisms that can produce such backgrounds.  Weak,
small-scale, and widespread absorption is likely to arise from turbulent
eddies, while strong, large-scale absorption appears organized in cloud
complexes along spiral arm shocks.  In the latter, the gas may be evolving from
an atomic to a molecular state prior to star formation, which would account for
the incomplete HISA-CO agreement.
\end{abstract}

\firstsection 
\section{Imaging the Cold Atomic Medium}

Cold atomic gas contains a large fraction of the mass of the interstellar
medium (ISM) and is a critical precursor to molecular cloud formation.
However, this ``cold atomic medium'' is hard to map in isolation, since warmer
gas is often brighter in traditional {\sc H~i} 21cm-line emission observations.
Fortunately, with proper angular resolution, cold atomic clouds can be imaged
as {\sc H~i} self-absorption (HISA) against warmer background {\sc H~i}
emission (\cite[Gibson et al. 2000]{g00}).  Recent large-scale radio synthesis
surveys like the Canadian and VLA Galactic plane surveys (CGPS: \cite[Taylor et
al. 2003]{cgps}; VGPS: \cite[Stil et al. 2006]{vgps}) are both well suited to
HISA studies.

\begin{figure}
\begin{center}
\begin{minipage}{4.0in}
\includegraphics[height=4.0in,width=4.0in,angle=0]{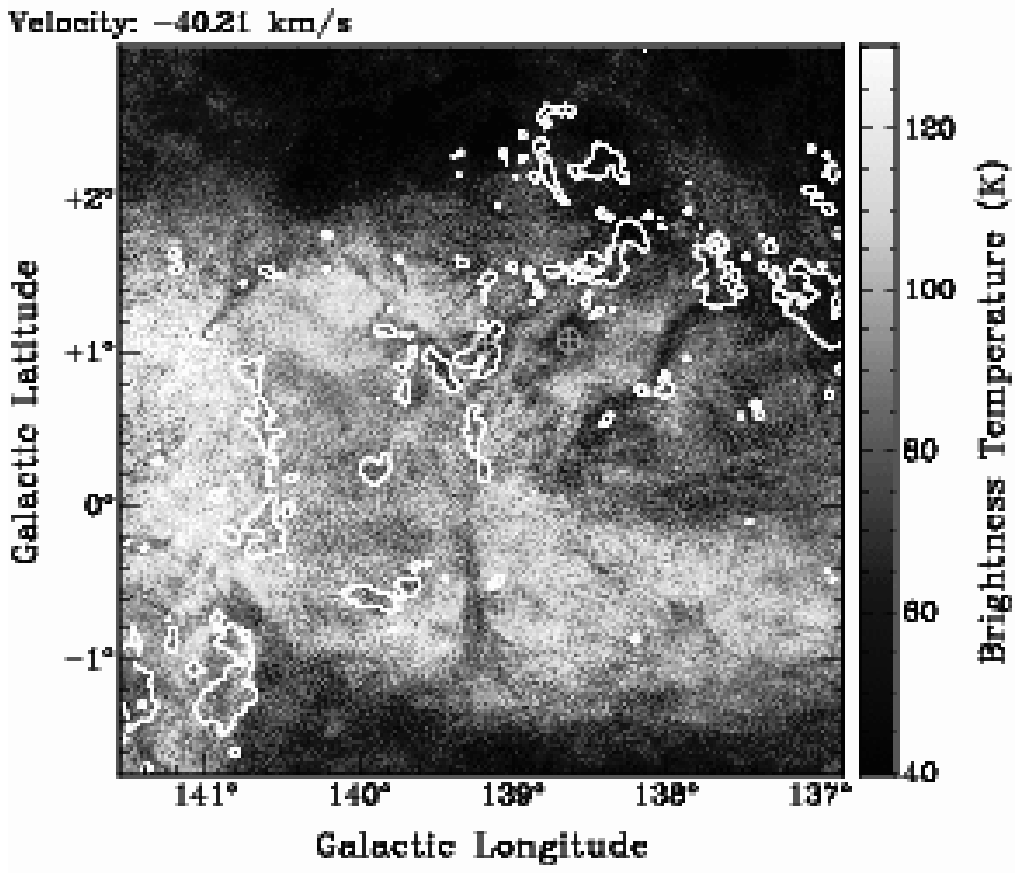}
\includegraphics[height=2.0in,width=2.0in,angle=0]{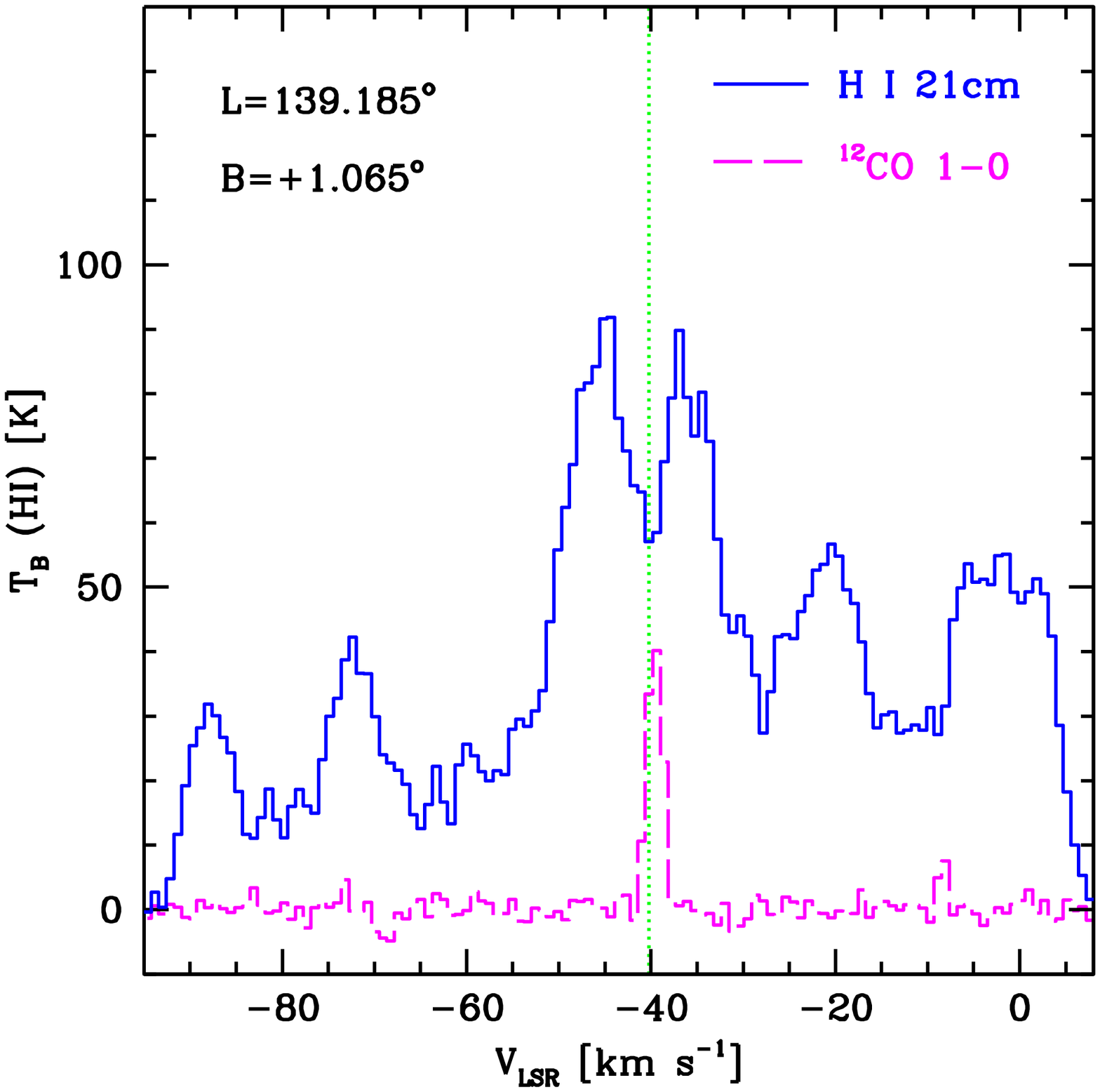}
\includegraphics[height=2.0in,width=2.0in,angle=0]{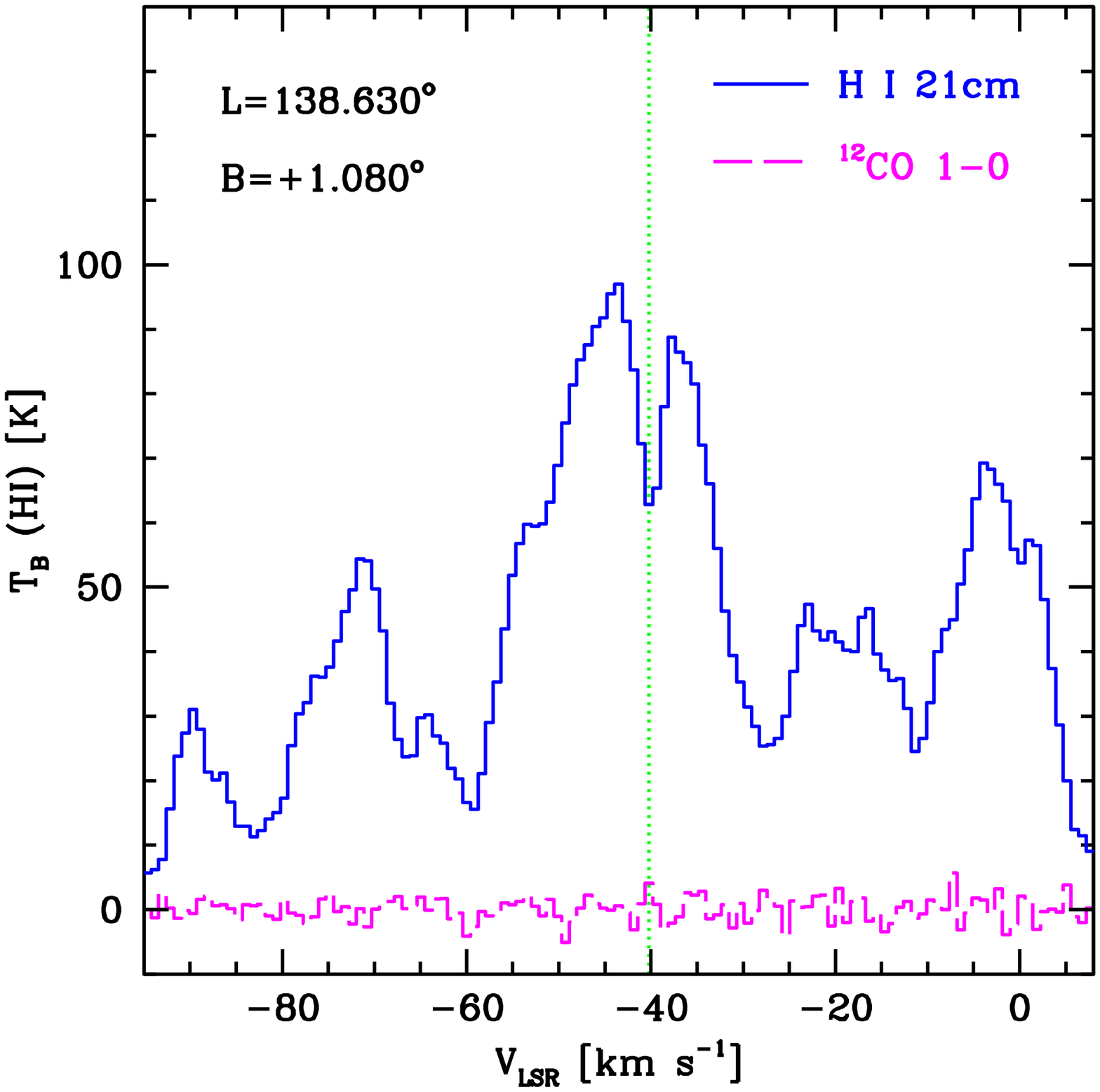}
\end{minipage}
\end{center}
\caption{
Sample HISA in the CGPS.  The upper panel shows an {\sc H~i} channel map with
$^{12}$CO 1-0 contours from \cite{ogs1}; the LSR velocity places this gas in
the Perseus spiral arm some 2 kpc away.  {\sc H~i} and CO spectra at the two
marked positions are plotted in the lower panels, showing that HISA is found
with and without CO, with the latter case more common in the outer Galaxy.
}\label{fig:hisa_example}
\end{figure}

{\bf Figure~\ref{fig:hisa_example}} compares some sample CGPS HISA to CO
emission.  The intricate structure of the HISA clouds is clear, as is their
frequent lack of apparent CO.  This lack is a puzzle unless significant H$_2$
is present without CO, since the low temperatures of many HISA features ($T <
50$~K) are hard to explain without molecular gas.  But if these clouds are
evolving rather than stable objects, then perhaps many have not yet formed
enough CO to detect (e.g., \cite[Klaassen et al. 2005]{kp05}).

Using an algorithm to identify and extract HISA features in the {\sc H~i} data,
we recently published a HISA census of the $73^\circ \times 9^\circ$ area
covered by the first 5 years of the CGPS (\cite[Gibson et al. 2005]{g05}).
As shown in {\bf Figure~\ref{fig:hisa_weak_strong}}, a low-level froth of weak,
disorganized HISA is found throughout the regions of the CGPS where the
background emission is bright enough for the HISA to be reliably detected.  By
contrast, stronger absorption is organized into discrete clouds and complexes.

\section{Velocity Perturbation Mechanisms}

HISA requires background line emission at the same radial velocity as the
foreground cloud.  Consequently, HISA radiative transfer probes both the
temperature and the velocity field of Galactic {\sc H~i}.  The CGPS is
primarily in the outer Galaxy, where pure differential rotation allows only one
position along the line of sight to have a particular velocity.  Since we
detect HISA in the outer Galaxy, the real velocity field must be perturbed from
this simple rotation model to provide background fields to absorb against.  Two
known mechanisms for this are turbulence and spiral density waves.

\begin{figure}
\begin{center}
\includegraphics[height=3.90625in,width=5.000in,angle=0]{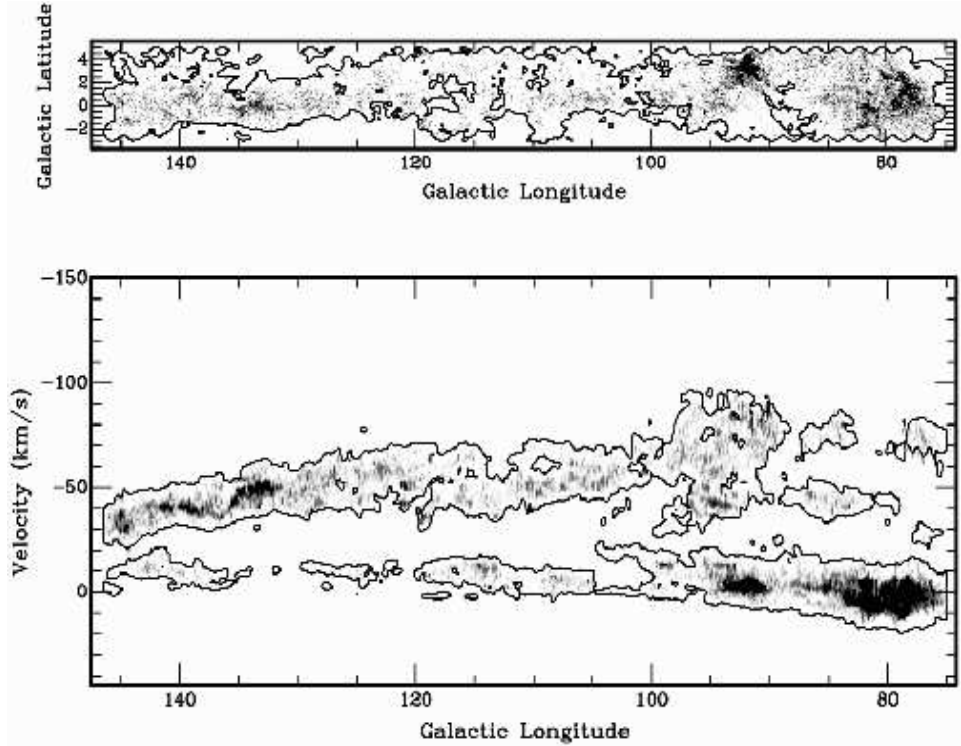}
\end{center}
\caption{
Longitude-latitude and longitude-velocity maps of HISA extracted from the
initial $73^\circ \times 9^\circ$ portion of the CGPS, in which the HISA ON-OFF
absorption amplitude is integrated over velocity (top) and latitude (bottom),
with stronger absorption being darker.  Contours mark where the {\sc H~i}
emission background becomes too faint for reliable HISA detection.
}\label{fig:hisa_weak_strong}
\end{figure}

The weak, widespread HISA in {\bf Figure~\ref{fig:hisa_weak_strong}} can be
explained as an ambient froth of cold atomic gas in the ISM made visible by
turbulent eddies; perhaps the cold gas is even a product of convergent
turbulent flows, as some models suggest (e.g., \cite[V\'{a}zquez-Semadeni et
al. 2006]{v06}).  The stronger HISA, however, is concentrated into distinct
complexes, especially along the Perseus arm near $-40\,{\rm km\,s^{-1}}$,
arguing for a more organized mechanism for these features.  The very cold
temperatures implied by this strong absorption could arise from gas that is
forming H$_2$ but has not yet formed much CO.

\section{Interpreting the Galactic HISA Distribution}

\begin{figure}
\begin{center}
\includegraphics[height=5.000in,width=2.5908in,angle=-90]{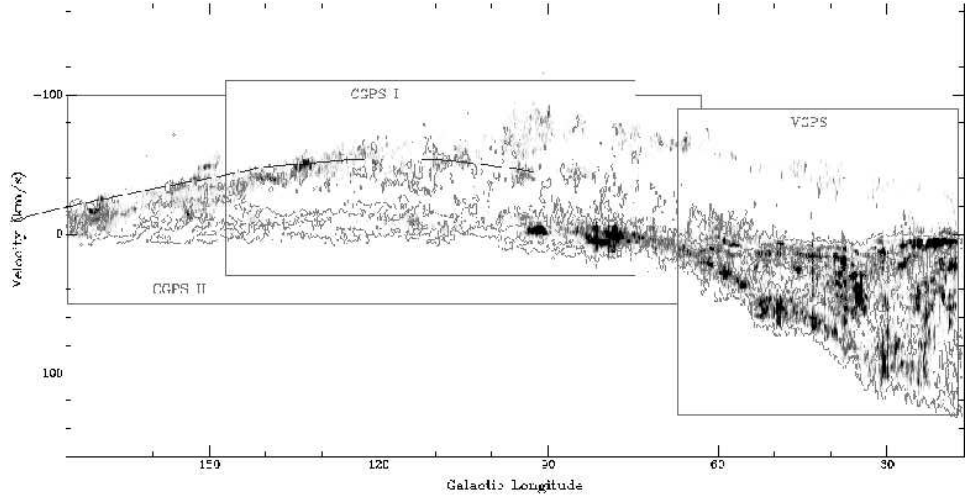}
\end{center}
\caption{
Longitude-velocity HISA distribution over the entire area covered by the
original CGPS, the CGPS II extension, and the VGPS.  Contours show molecular
gas traced by \cite{dht01} $^{12}$CO 1-0 emission, and the curved line marks
the maximum velocity departure from Galactic rotation predicted by the Perseus
arm spiral shock model of \cite{r72}.
}\label{fig:hisa_cvgps}
\end{figure}

{\bf Figure~\ref{fig:hisa_cvgps}} shows a new HISA survey incorporating the
VGPS and extensions to the CGPS (\cite[Gibson et al. 2006]{g06}).  This plot
includes CO contours and a curve marking the maximum velocity departure from
circular rotation predicted by \cite{r72} for the Perseus spiral shock.  Apart
from some irregular ISM structure Roberts did not model, the strong Perseus
HISA lies near the shock curve but at less extreme velocities, consistent with
clouds lying just downstream of the spiral shock.  Similar but fainter
shock-related HISA may be seen in the Outer arm at negative velocities in the
VGPS data.  Both arms have a poor HISA-CO match, as would be expected for
evolving gas; in this scenario, the visible CO traces more evolved gas further
downstream, where little background {\sc H~i} emission is left to show the
remaining cold atomic gas in the CO clouds as HISA.

In the inner Galaxy, the picture is more complex.  A much larger amount of HISA
is seen, since even simple rotation provides near and far points on each sight
line with the same velocity.  This allows a much more widespread {\sc H~i}
emission background, which may account for the stronger HISA-CO agreement and
the lack of clearly-defined spiral arms.  However, the latter requires a
significant amount of cold interarm {\sc H~i} in the disk.  Whether the
interarm cold cloud population is related to the outer-Galaxy turbulent HISA
population is still under investigation.

\begin{acknowledgments}
This work has been supported by the National Astronomy and Ionosphere
Center operated by Cornell University under Cooperative Agreement with the
U.S. National Science Foundation and by grants from the Natural Sciences and
Engineering Research Council of Canada to the University of Calgary.
\end{acknowledgments}

%
%

\end{document}